\documentclass[preprints,article,accept,moreauthors,pdftex]{Definitions/mdpi}
\usepackage{amssymb}
\usepackage{braket}

\DeclareMathOperator{\trace}{tr} % trace

\firstpage{1}
\pubvolume{1}
\issuenum{1}
\articlenumber{0}
\pubyear{2021}
\copyrightyear{2020}
\externaleditor{}
\datereceived{}
\dateaccepted{}
\datepublished{}
\hreflink{http://doi.org/}
\Title{Casimir Interaction Between a Plane and a Sphere: Correction to the Proximity-Force 
Approximation at Intermediate Temperatures}
\TitleCitation{Casimir Interaction Between a Plane and a Sphere: Correction to the Proximity-Force 
Approximation at Intermediate Temperatures}

\Author{Vinicius Henning $^{1}$\orcidA{}, Benjamin Spreng $^{2,3}$\orcidB{},
Paulo A. Maia Neto $^{1}$\orcidC{} and Gert-Ludwig Ingold $^{2}$\orcidD{}}

\AuthorNames{Vinicius Henning, Benjamin Spreng, Paula A. Maia Neto and Gert-Ludwig Ingold}

\AuthorCitation{Henning, V.; Spreng, B.; Maia Neto, P. A.; Ingold, G.-L.}

\address{$^{1}$ \quad Instituto de F{\'i}sica, Universidade Federal do Rio de Janeiro, CP 68528, Rio de Janeiro RJ 21941-909, Brazil\\
$^{2}$ \quad Universit{\"a}t Augsburg, Institut f{\"u}r Physik, 86135 Augsburg, Germany\\
$^{3}$ \quad Department of Electrical and Computer Engineering, University of California, Davis, California 95616, USA}

\corres{Correspondence: gert.ingold@physik.uni-augsburg.de (G.I.)}

\abstract{We consider the Casimir interaction energy between a plane and a
sphere of radius $R$ at finite temperature $T$ as a function of the distance of
closest approach $L$.  Typical experimental conditions are such that the
thermal wavelength $\lambda_T=\hbar c/k_\mathrm{B}T$ satisfies the condition $L\ll
\lambda_T\ll R$.  We derive the leading correction to the proximity-force
approximation valid for such intermediate temperatures by developing the
scattering formula in the plane-wave basis.  Our analytical result
captures the joint effect of the spherical geometry and temperature and is
written as a sum of temperature-dependent logarithmic terms.
Surprisingly, two of the logarithmic terms arise from the Matsubara
zero-frequency contribution.}


\begin{document}

\section{Introduction}
\label{sec:introduction}

The Casimir effect is a striking consequence of the zero-point energy of
the quantum electromagnetic field. The geometry  studied by Casimir
himself was the interaction between two planar perfectly-reflecting plates in vacuum,
which experience an attractive force \cite{Casimir1948,Bordag2009}. However,
due to parallelism issues, most experiments are performed using either a plane-sphere \cite{Sushkov2011,Torricelli2011,
Chang2012, GarciaSanchez2012, Banishev2013, Sedighi2016, Bimonte2016,LeCunuder2018} or a sphere-sphere \cite{Wodka2014,Ether2015,Garrett2018} geometry
(for reviews see~\cite{Klimchitskaya2009,Decca2011,Lamoreaux2011,Klimchitskaya2020,Gong2021}).

In contrast to the well understood
 plane-plane geometry, an exact theoretical formalism
for the plane-sphere \cite{Emig2008,MaiaNeto2008} and the sphere-sphere~\cite{Emig2007}
geometries became available only with the advent of the
scattering approach~\cite{Lambrecht2006,Rahi2009}. However,
experimental data for the Casimir force continued to be compared with theoretical results obtained within
the proximity-force  approximation (PFA) due to Derjaguin~\cite{Derjaguin1934}
as numerical implementations of the scattering formula for experimentally relevant geometrical aspect ratios 
were not available until very recently \cite{Hartmann2017,Hartmann2018,Hartmann2020,Spreng2020}.
Within the PFA, the Casimir energy is obtained from the Lifshitz's formula for parallel planes \cite{Lifshitz1956,Kats1977,Jaekel1991,Genet2003} by averaging over 
the local surface-to-surface distance~\cite{Parsegian2006}.

Starting from the exact scattering approach for spherical surfaces, the PFA result was obtained~\cite{Spreng2018} as the leading 
 asymptotics for large sphere radius $R$ from a WKB saddle-point contribution \cite{Nussenzveig69,Khare1975,Nussenzveig1992,Grandy2005}.
 The saddle point has a direct physical interpretation in terms of
 specular reflection at the points of closest approach on each interacting surface~\cite{Spreng2018}. 
Carrying out the semiclassical approximation up to the next-to-leading order, 
the zero-temperature leading order correction to PFA  \cite{Teo2011,Bimonte2012EPL}
was re-derived and shown to be mostly due to corrections to the WKB approximation for the Mie scattering amplitudes~\cite{Henning2019}. 
 
Investigations of the leading-order correction to PFA became particularly
relevant on account of recent experiments
\cite{Krause2007,Garrett2018,Liu2019,Liu2019B} probing larger aspect ratios
$L/R$ where $L$ represents the distance of closest approach.  In those
experiments, thermal effects have to be considered since the contribution from
thermal photons becomes more important as the distance $L$ is increased
\cite{Sauer1962,Mehra1967,Bostrom2000,Genet2000,Ingold2009},
especially when modelling experiments with colloidal suspensions
\cite{Wodka2014,Ether2015} with a near index matching at non-zero Matsubara
frequencies \cite{Parsegian1970}.  Thus, a theoretical approach taking into
account both thermal and beyond-PFA geometrical effects is required in most
cases where a measurable deviation from PFA is expected. 

In this paper, we derive the analytical leading-order correction to the PFA
result for intermediate temperatures satisfying  $L\ll \lambda_T\ll R,$ where
$\lambda_T=\hbar c/k_\mathrm{B}T$ is the thermal wavelength.  Such condition holds in
typical Casimir experiments as $\lambda_T\approx 7.6\,\mu{\rm m}$ at
$T=300\,{\rm K}$. We consider the plane-sphere setup within  the
perfectly-reflecting model for simplicity. However, our method based on a
semiclassical expansion developed in the plane-wave
basis~\cite{Spreng2018,Henning2019} can also be applied to the sphere-sphere
geometry and to real materials. 

The non-trivial interplay between geometrical and thermal corrections was numerically demonstrated for a scalar field model within the
worldline approach~\cite{Weber2010,Weber2010B}.
Finite-temperature numerical implementations of the scattering approach based either on spherical multipoles \cite{Hartmann2017, Hartmann2018,
Canaguier-Durand2010,
Canaguier-Durand2010B,Zandi2010,Rodriguez-Lopez2011,Umrath2016}
 or plane waves \cite{Spreng2020} provided further evidence 
that the thermal and curvature effects are strongly correlated. 
The high-temperature limit is amenable to analytical~\cite{Bimonte2012} and numerical~\cite{Bimonte2017, Bimonte2018b} calculations based on bispherical coordinates. 
In the case of perfect reflectors, the leading-order corrections to PFA for low temperature \cite{Bordag2010temp}, $L\ll R \ll \lambda_T$, 
and high temperatures \cite{Bimonte2017}, $\lambda_T \ll L\ll R$, were derived analytically
 by considering the asymptotic limit of the scattering matrices in the
 multipolar spherical basis.

The derivative expansion provides yet another method to obtain the leading-order correction to PFA~\cite{Bimonte2012EPL,Fosco2011}.
It relies on a re-summation of the perturbative expansion around the parallel-planes geometry~\cite{Fosco2014}.
The derivative expansion 
is implemented by approximating 
the perturbative kernel by its power series up to second order in the momentum variable $k$. 
 For a scalar field satisfying Neumann boundary conditions, the finite-temperature kernel is not analytical at $k=0$
  in the case of three spatial dimensions, and then 
the derivative expansion breaks down~\cite{Fosco2012DE}. 
This is also the case for the electromagnetic Casimir effect in three dimensions when considering perfectly-reflecting or plasma mirrors \cite{Fosco2015}.
The singular behavior of the perturbative kernel indicates that the correction to PFA is of a nonlocal nature at finite temperatures. 

Such non-analytical and nonlocal behavior translates into a correction to PFA containing powers of $\log(L/R),$ 
as first discussed in connection with the high-temperature regime~\cite{Canaguier-Durand2012}.
We show that Bimonte's  $\log^2(L/R)$ leading correction arising from the Matsubara zero-frequency contribution~\cite{Bimonte2017}, which is usually associated 
to the high-temperature regime, should also be kept when $L\ll \lambda_T\ll R$. 
By developing the scattering formula in the plane-wave basis, we 
 re-derive Bimonte's result as well as the next-to-leading order
correction proportional to $\log(L/R).$ The latter turns out to be also required for an accurate description of experimentally-relevant aspect ratios. 
For the contribution of the non-zero Matsubara frequencies, 
we derive a correction proportional to $\log^2(L/\lambda_T)$
by employing the Euler-Maclaurin sum formula. 

The paper is organized as follows. Section~\ref{sec:energy_pwb} presents in a
first part the basic tools and notations required to expand the scattering
formula in the plane-wave basis. A second part discusses the asymptotic
expansion in powers of the inverse sphere radius and introduces general
expressions for the leading-order correction to PFA.
Section~\ref{sec:LO-Matsubara} is devoted to an explicit evaluation of the
leading-order correction for individual Matsubara frequencies. A particular
focus will be put on the special case of the zero Matsubara frequency. The
results from this section will be used in Sec.~\ref{sec:asym_expansion} to
derive the leading-order correction to PFA valid for intermediate temperatures.
In the analysis, we will distinguish between the contributions arising from the
geometric optical WKB approximation and from its diffraction correction.
Concluding remarks are presented in Sec.~\ref{sec:conclusions}. A review of
the next-to-leading term for the saddle-point approximation for the one-dimensional
case together with the results for the multidimensional generalization is given
in Appendix~\ref{app:SP_formula}.

\section{Asymptotic expansion of the Casimir free energy in the plane-wave basis}
\label{sec:energy_pwb}

\subsection{Casimir free energy for plane-sphere geometry}

\begin{figure}
 \begin{center}
  \includegraphics[scale=1]{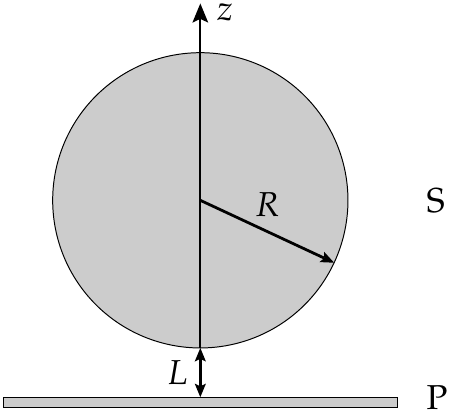}
 \end{center}
	\caption{Sphere (S) of radius $R$ and plate (P) separated by a distance $L$.}
 \label{fig:geometry}
\end{figure}

We consider a spherical surface of radius $R$ close to a plate as illustrated in
Fig.~\ref{fig:geometry} and assume both surfaces to be perfectly reflecting.
The plate lies in the $xy$--plane and the $z$-axis perpendicular to it goes
through the sphere center. The closest distance between plate and sphere is
denoted by $L$.

Within the scattering approach~\cite{Lambrecht2006}, the Casimir free energy is written as a sum over the Matsubara frequencies $\xi_n=2\pi nk_\mathrm{B}T/\hbar$
\begin{equation}
\label{eq:Matsubara_sum}
	 \mathcal{F} = \sum_{n=0}^{\infty}{}^{'}  \mathcal{F}_n
\end{equation}
with
\begin{equation}
\label{eq:casimir_energy}		
	 \mathcal{F}_n = k_\mathrm{B}T\trace\log\big[1-\mathcal{M}(\xi_n)\big]
\end{equation}
and where the prime indicates that the term $n=0$ is multiplied by one-half. 
The operator $\mathcal{M}(\xi_n)$ describes the round trip of an electromagnetic wave in the empty gap between the
two interacting surfaces. 
For the geometry shown in
Fig.~\ref{fig:geometry}, the round-trip operator is decomposed as
\begin{equation}
 \label{eq:round_trip_operator}
 \mathcal{M}(\xi_n) = \mathcal{T}_\text{PS} \mathcal{R}_\text{S}
	       \mathcal{T}_\text{SP} \mathcal{R}_\text{P}\,.
\end{equation}
$\mathcal{R}_\text{S}$ and $\mathcal{R}_\text{P}$ are the reflection operators
for sphere and plate taken with respect to reference points at the sphere
center and at the intersection between the $z$-axis and the plate surface,
respectively. $\mathcal{T}_\text{PS}$ describes the translation from the first
to the second reference point over a distance $L+R$ along the $z$-axis, while
$\mathcal{T}_\text{SP}$ accounts for the translation in the opposite direction.
 
We expand the logarithm in (\ref{eq:casimir_energy}) in powers of the round-trip operator $\mathcal{M}(\xi_n)$:
\begin{equation}\label{eq:energy_round_trips}
	\mathcal{F}_n = - k_\mathrm{B}T
		\sum_{r=1}^{\infty}\frac{1}{r}\trace\mathcal{M}(\xi_n)^r \,.
\end{equation}
The summation variable $r$ in (\ref{eq:energy_round_trips}) represents the
number of round trips between the two interacting surfaces. Thus, the Casimir
free energy collects all the contributions from one to infinitely many round
trips within the empty cavity bounded by the reflecting surfaces.

We evaluate the trace in (\ref{eq:energy_round_trips}) in the plane-wave basis
$\{|\mathbf{k}, \phi, p\rangle\}$ as defined by the angular spectral
representation \cite{Nieto-Vesperinas2006}. Here, ${\bf k}$ denotes the
projection of the wave vector onto the $xy$--plane, $\phi=\pm 1$ defines the
sense of propagation along the $z$-axis in upwards or downwards direction,
respectively, and the polarization $p$ is either transverse electric (TE) or transverse
magnetic (TM). The projected wave vector ${\bf k}$ and the Matsubara
frequency $\xi_n$ jointly define the Wick-rotated axial component of the three-dimensional
wave vector
\begin{equation}
 \label{eq:imag_dispersion}
 \kappa = \sqrt{\frac{\xi_n^2}{c^2}+\mathbf{k}^2}\,.
\end{equation}

The translation operators $\mathcal{T}_\text{PS}$ and $\mathcal{T}_\text{SP}$
are diagonal in the plane-wave basis with eigenvalues $e^{-\kappa (R+L)}$.
The action of the reflection operator at the planar surface
\begin{equation}
\mathcal{R}_\text{P}|\mathbf{k},-,p\rangle = r_p
		|\mathbf{k},+,p\rangle\,,
\end{equation}
conserves the projected wave vector ${\bf k}$ as well as the polarization $p$. Here,
$r_p$ are the standard Fresnel coefficients for specular reflection which, for
the case of perfect reflectors, are given by  $r_\text{TM} = 1$ and
$r_\text{TE}=-1$.

In contrast, ${\bf k}$ and $p$ are not conserved during a reflection at the
spherical surface.  The contribution corresponding to $r$ round-trips in
(\ref{eq:energy_round_trips}) apart from the trace thus implies an integration
over $r-1$ intermediate wave vectors $\mathbf{k}_{1},\ldots,\mathbf{k}_{r-1}$
and a summation over intermediate polarizations $p_1,\ldots,p_{r-1}$ taking
values TE or TM
\begin{equation}
 \label{eq:trMr}
 \trace\mathcal{M}(\xi_n)^r = \sum_{p_0,\dots,p_{r-1}}
   \int\prod_{j=0}^{r-1}
   \frac{\mathrm{d}\mathbf{k}_j}{(2\pi)^2} e^{-2\kappa_j(L+R)}r_{p_j}
   \langle\mathbf{k}_{j+1},-,p_{j+1}|\mathcal{R}_\mathrm{S}|\mathbf{k}_j,+,p_j\rangle\,.
\end{equation}
We use a cyclic index convention such that $j=r$ is equivalent to $j=0$.  The
matrix elements of the reflection operator $\mathcal{R}_\text{S}$ appearing in
(\ref{eq:trMr}) can be written in terms of the standard Mie scattering
amplitudes together with coefficients describing the change between the Fresnel
and the scattering polarization basis \cite{Spreng2018,Henning2019}. For the evaluation
of the leading-order (LO) PFA result and its LO correction for the perfectly reflectors
case, the relevant matrix elements effectively reduce to \cite{Henning2019}
\begin{equation}
\label{eq:RS}
 \begin{aligned}
 \braket{\mathbf k_j, -, \mathrm{TM} | \mathcal{R}_\mathrm{S} | \mathbf k_i, +, \mathrm{TM}} &= \frac{2\pi c}{\xi \kappa_j} S_2\\
 \braket{\mathbf k_j, -, \mathrm{TE} | \mathcal{R}_\mathrm{S} | \mathbf k_i, +, \mathrm{TE}} &= \frac{2\pi c}{\xi \kappa_j} S_1
 \end{aligned}
\end{equation}
while the matrix elements involving the coupling between different polarizations do not contribute,
\textit{i.e.},  $\braket{\mathbf k_j, \mathrm{TM} | \mathcal{R}_\mathrm{S} | \mathbf k_i, \mathrm{TE}}
=  \braket{\mathbf k_j, \mathrm{TE} | \mathcal{R}_\mathrm{S} | \mathbf k_i, \mathrm{TM}} = 0$.
The Mie scattering amplitudes $S_1$ and $S_2$ \cite{BH} in \eqref{eq:RS} are functions of the
imaginary size parameter $\xi R/c$ and the scattering angle $\Theta$ defined through
\begin{equation}
\cos(\Theta) = -\frac{c^2}{\xi^2}(\kappa_i \kappa_j + \mathbf{k}_i \cdot \mathbf{k}_j)\,.
\end{equation}

For large spheres, the Mie scattering amplitudes can be expanded as 
\cite{Nussenzveig69,Khare1975,Nussenzveig1992,Grandy2005}
\begin{equation}
 \label{eq:SA_with_corrections}
 S_p = S_p^\text{WKB}\left(1 + \frac{1}{R}s_p + \mathcal{O}\left(R^{-2}\right)\right) \,.
\end{equation}
with the leading-order contribution given by the WKB expression
\begin{equation}
\label{eq:S_WKB}
S_p^\text{WKB} = (-1)^p\frac{\xi R}{2c} \exp\left[\frac{2\xi R}{c} \sin\left(\frac{\Theta}{2}\right)\right]
\end{equation}
and
\begin{align}
s_1 &= \frac{c}{2\xi} \frac{\cos(\Theta)}{\sin^3(\Theta/2)} =
  - \frac{\sqrt{2}(\kappa_i \kappa_j + \mathbf{k}_i \cdot \mathbf{k}_j)}
	 {(\xi^2/c^2 + \kappa_i \kappa_j + \mathbf{k}_i \cdot \mathbf{k}_j)^{3/2}}
  \label{eq:s1corr}\\
s_2 &= -\frac{c}{2\xi} \frac{1}{\sin^3(\Theta/2)} =
  -\frac{\sqrt{2}\xi^2/c^2}{(\xi^2/c^2 + \kappa_i \kappa_j + \mathbf{k}_i \cdot \mathbf{k}_j)^{3/2}}\,.
  \label{eq:s2corr}
\end{align}
describing the leading-order corrections.

\subsection{Asymptotic expansion}

Evaluating the trace \eqref{eq:trMr} within the lowest-order saddle-point
approximation (LO-SPA) together with the WKB expression \eqref{eq:S_WKB} for the Mie
scattering amplitudes, one obtains by means of \eqref{eq:Matsubara_sum} and
\eqref{eq:energy_round_trips} the Casimir free energy within the
proximity-force approximation \cite{Spreng2018}. This result constitutes the
leading term in an asymptotic expansion for large sphere radius $R$ and can be
entirely understood in terms of geometrical optics.

Our aim is to go beyond the proximity-force approximation and to determine the
corrections which are smaller by a factor $1/R$. Two corrections need to be
taken into account. Firstly, in the evaluation of the trace \eqref{eq:trMr} one
needs to go one order beyond the LO-SPA. We refer to this correction as
next-to-leading order saddle-point approximation (NTLO-SPA). Since this
correction is not as widely known as the LO-SPA, we give some details in
Appendix~\ref{app:SP_formula}. In the evaluation of the NTLO-SPA, the Mie
scattering amplitudes are still to be taken within the WKB approximation and we
are thus still within the realm of geometrical optics. A second contribution to
the correction to the proximity-force approximation arises from the leading
correction to the WKB Mie scattering amplitudes as specified by
\eqref{eq:SA_with_corrections} together with \eqref{eq:s1corr} and
\eqref{eq:s2corr}. This contribution takes diffraction into account. For this
second contribution, it is sufficient to evaluate the integrals in
\eqref{eq:trMr} within LO-SPA.

Inserting (\ref{eq:SA_with_corrections}) in (\ref{eq:trMr}), allows us to
express the trace over the $r$-th round trip in the form
\begin{equation}
\label{eq:rloops}
\trace\mathcal{M}(\xi_n)^r \simeq \left(\frac{R}{4\pi}\right)^r\int {\mathrm{d}}\mathbf{k}_0 \dots \mathrm{d}\mathbf{k}_{r-1}\,g(\mathbf{k}_0,\dots,\mathbf{k}_{r-1}) e^{-R f(\mathbf{k}_0,\dots,\mathbf{k}_{r-1})}
\end{equation}
with
\begin{equation}
 \label{eq:g_definition}
 g(\mathbf{k}_0,\dots,\mathbf{k}_{r-1}) = \sum_{p=1,2}\prod_{j=0}^{r-1}
 \frac{e^{-2\kappa_jL}}{\kappa_j}\left(1 + \frac{1}{R}s_p\right)
\end{equation}
and
\begin{equation}
 \label{eq:f_definition}
 f(\mathbf{k}_0,\dots,\mathbf{k}_{r-1}) = \sum_{j = 0}^{r-1} \left(\kappa_j + \kappa_{j+1} - \left[2\left(\xi_n^2/c^2 +
	\kappa_j\kappa_{j+1}+\mathbf{k}_{j}\cdot\mathbf{k}_{j+1} \right)\right]^{1/2}\right) \,.
\end{equation}
Note that $s_p$ in \eqref{eq:g_definition} depends on the indices $j$ and $j+1$
through the respective wave vectors.

The $2r$-dimensional integral in \eqref{eq:rloops} is suitable for a saddle-point approximation
where $R$ plays the role of the large parameter. It is straightforward to show that there exists
a continuous family of saddle points
\begin{equation}
\label{eq:SP_manifold}
\mathbf{k}_0 = \dots = \mathbf{k}_{r-1} \equiv \mathbf{k}_\mathrm{sp}
\end{equation}
parameterized by $\mathbf{k}_\mathrm{sp}$. While the saddle-point approximation
can be applied in the directions orthogonal to the saddle-point manifold, in the end we will
be left with an integral over the saddle-point manifold which needs to be evaluated exactly.

As a consequence of the existence of a continuous family of saddle points, the
Hessian matrix of \eqref{eq:f_definition} is singular with two vanishing
eigenvalues in view of the two-dimensional character of
$\mathbf{k}_\mathrm{sp}$. In order to cope with the vanishing eigenvalues, it
is convenient to transform the Hessian matrix into its eigenbasis as described
in Ref.~\cite{Henning2019}. For completeness, we review in the following the basic steps.

On the saddle-point manifold \eqref{eq:SP_manifold}, the Hessian matrix can be brought into
block-diagonal form 
\begin{equation}\label{eq:hessian}
\mathsf{H}= \begin{pmatrix}
\mathsf{H}_{xx} & 0 \\
0               & \mathsf{H}_{yy}
\end{pmatrix}
\end{equation}
with the matrix elements
\begin{equation}
 \label{eq:hessian_xx}
 \big(\mathsf{H}_{xx}\big)_{ij} = \left.\frac{\partial^2f}{\partial k_{i,x}\partial k_{j,x}}
				  \right\vert_{\mathrm{sp}}
\end{equation}
and a corresponding expression for $\mathsf{H}_{yy}$.

The blocks of the Hessian matrix can be expressed as
$\mathsf{H}_{xx} = \mathsf{H}_{yy} = (1/2 \kappa_\mathrm{sp})\Gamma_r$ in terms of
the $r\times r$ circulant matrix
\begin{equation}
\label{eq:Gamma1}
\Gamma_r = \begin{pmatrix}
2 & -1 &&& -1\\
-1 & 2 & -1 && \\
& -1 & \ddots & \ddots &\\
& & \ddots & \ddots & -1\\
-1& & & -1 & 2 \\
\end{pmatrix}
\end{equation}
for $r\ge 3$ and where the matrix elements not shown are zero. In the case of two round trips
\begin{equation}
\Gamma_2 = \begin{pmatrix}
2 & -2 \\
-2 & 2
\end{pmatrix}
\end{equation}
(note that the corresponding expression in \cite{Henning2019} is missing a factor of 2)
while for $r=1$ $f\equiv0$.

It is now convenient to introduce transformed variables $v$ through
\begin{equation}
 \label{eq:transformation}
 k_{j,x} = \sum_{l=0}^{r-1}\mathsf{W}_{jl} v_{l,x}
\end{equation}
with the Fourier matrix
\begin{equation}
\label{eq:transformation_matrix}
 \mathsf{W}_{jl}= \frac{1}{\sqrt{r}}\exp\left(\frac{2\pi i}{r}jl\right)
\end{equation}
and correspondingly for the $y$-direction.

After the transformation, the two blocks of the Hessian matrix are of counter-diagonal form
\begin{equation}
 \label{eq:WHW}
 \big(\mathsf{W}^T\mathsf{H}_{xx}\mathsf{W}\big)_{jl} = \lambda_j\delta_{j,r-l}
\end{equation}
with the eigenvalues
\begin{equation}
 \label{eq:eigenvalues}
 \lambda_j = \frac{2}{\kappa_\text{sp}}\sin^2\left(\frac{\pi j}{r}\right)
\end{equation}
and $j=0, 1,\ldots, r-1$. As expected, one eigenvalue ($j=0$) vanishes for each block
and the variables $v_{0,x}$ and $v_{0,y}$ parametrize the two-dimensional saddle-point manifold.

Applying the saddle-point approximation \eqref{eq:SP_formula} with
\eqref{eq:LO-SPA_multi} and \eqref{eq:NTLO-SPA_multi}, \eqref{eq:rloops} can
now be expressed as \cite{Henning2019}
\begin{equation}
 \label{eq:asym_formula}
 \trace\mathcal{M}(\xi_n)^r = \frac{R}{2r}\int_{\xi_n/c}^\infty\mathrm{d}\kappa_\mathrm{sp} \, \kappa_\mathrm{sp}^r
	\left[F_0+\frac{1}{R}F_1+o\left(R^{-1}\right)\right]\,,
\end{equation}
where we have transformed the variables $v_{0,x}$ and $v_{0,y}$ back to the
wave vector at the saddle point. The first and second terms in the integrand 
correspond to the LO-SPA and NTLO-SPA, respectively, and
are given by
\begin{equation}
 \label{eq:F0}
 F_0 = g\vert_\mathrm{sp}
\end{equation}
and
\begin{equation}
 \label{eq:F1}
 F_1 = g\vert_\mathrm{sp}\left(\sum_{ijk}\frac{f_{ijk}f_{\bar i\,\bar j\,\bar k}}{12\lambda_i\lambda_j\lambda_k}
			  -\sum_{ij}\frac{f_{i\bar ij\bar j}}{8\lambda_i\lambda_j}\right)
              +\sum_i\frac{g_{i\bar i}}{2\lambda_i}
\end{equation}
where we have introduced the shorthand notation $\bar{i}=r-i$. The summation
runs over the indices from $1$ to $r-1$ and implies also a summation over the
corresponding components $x$ and $y$. The indices at the functions $f$ and $g$
denote derivatives with respect to the corresponding components of the variables
$v$ evaluated at the saddle point. Note that in comparison with \eqref{eq:NTLO-SPA_multi}
the second and the fourth term are missing which were shown in Ref.~\cite{Henning2019}
not to contribute to \eqref{eq:F1}.

\section{Leading-order correction for individual Matsubara frequencies}
\label{sec:LO-Matsubara}

For an asymptotic expansion in powers of the inverse sphere radius, the radius
$R$ has to be compared with the other length scales of the problem. While the
radius can be chosen larger than $c/\xi_n$ for all $n > 0$, the zero Matsubara
frequency $\xi_0$ may require special care.  Interestingly, it turns out that
the leading order terms for the matrix elements \eqref{eq:RS} and as a
consequence for the trace \eqref{eq:trMr} hold for all Matsubara frequencies.
The expressions for the zero Matsubara frequency can thus be obtained by taking
the zero-frequency limit of the results for positive Matsubara frequencies.
Therefore, the PFA result holds for arbitrary temperatures including the
high-temperature limit determined by the contribution of the zero Matsubara
frequency \cite{Spreng2018}.

The situation is different when the next-to-leading order term is considered.
In contrast to the contributions due to positive Matsubara frequencies, the
zero-frequency contribution cannot be obtained from the known diffraction
correction \cite{Nussenzveig69,Khare1975,Nussenzveig1992,Grandy2005} to the WKB
Mie scattering amplitudes \eqref{eq:S_WKB}. Proceeding on that basis would
yield an infrared divergence in the corresponding integral
\eqref{eq:asym_formula} over $k_\mathrm{sp}$.

We will start by discussing the case of positive Matsubara frequencies in
Sec.~\ref{sec:positive_Matsubaras} where we make use of results obtained
earlier in Ref.~\cite{Henning2019}. In Sec.~\ref{sec:Matsubara_n0}, we will
then derive the asymptotic expansion of the zero-frequency contribution to
obtain both the NTLO and the next-to-next-to-leading order (NNTLO) terms.  The
latter turns out to be non-negligible for experimentally relevant aspect ratios
and then should be kept alongside the former, which was first derived in
Ref.~\cite{Bimonte2017} by the multipolar approach. In
Sec.~\ref{sec:Matsubara_n0}, we focus on the TE zero-frequency contribution, as
the TM correction can be more easily derived from an exact analytical
representation obtained either by using bispherical coordinates
\cite{Bimonte2012} or by developing the plane-wave basis representation
(\ref{eq:trMr}) \cite{Schoger2020}.

\subsection{Positive Matsubara frequencies}
\label{sec:positive_Matsubaras}

We first turn to the discussion of Matsubara frequencies $\xi_n$ with $n>0$
and consider the two contributions to the integrand in \eqref{eq:asym_formula}.
The leading-order term \eqref{eq:F0} is obtained by evaluating \eqref{eq:g_definition}
at the saddle point and can be decomposed into contributions from the two polarizations
as
\begin{equation}
F_0 = g_\mathrm{TE} + g_\mathrm{TM}
\end{equation}
with
\begin{equation}
\begin{aligned}
g_\mathrm{TE} &= \frac{e^{-2r\kappa_\mathrm{sp} L}}{\kappa_\mathrm{sp}^r}\left(1 + \frac{r(\xi_n^2 - 2c^2 \kappa_\mathrm{sp}^2)}{2c^2\kappa_\mathrm{sp}^3 R} \right)\\
g_\mathrm{TM} &= \frac{e^{-2r\kappa_\mathrm{sp} L}}{\kappa_\mathrm{sp}^r}\left(1 - \frac{r\xi_n^2}{2c^2\kappa_\mathrm{sp}^3 R}\right)\,.
\end{aligned}
\end{equation}

The evaluation of the next-order term \eqref{eq:F1} is more involved. We refer the reader
to appendix A in \cite{Henning2019} for details. There, it was found that
\begin{equation}
F_1 = - \frac{(r^2-1)\left(rL\kappa_\mathrm{sp}(c^2\kappa_\mathrm{sp}^2+\xi_n^2)+ \xi_n^2\right)}
	{6rc^2\kappa_\mathrm{sp}^3} \frac{e^{-2r\kappa_\mathrm{sp} L}}{\kappa_\mathrm{sp}^r}\,.
\end{equation}

The leading term in the $1/R$ expansion corresponding to the PFA result is
determined entirely by local scattering channels describing specular reflection
at the point of closest approach on the spherical surface. It can thus be
completely understood in terms of geometrical optics. In contrast, the NTLO
term consists of two contributions
\begin{equation}
\label{eq:tr-ntlo}
[\trace\mathcal{M}(\xi_n)^r]_\mathrm{\scriptscriptstyle {NTLO}}
= \sum_{p=\mathrm{TE,TM}}\left([\trace\mathcal{M}(\xi_n)^r]^p_\mathrm{d} + 
		[\trace\mathcal{M}(\xi_n)^r]^p_\mathrm{go}\right)
\end{equation}
of different physical origin. The first term carrying the subscript ``d''
captures the effect of diffraction as it arises from the LO correction to the
WKB approximation for the Mie scattering amplitudes taken at the LO-SPA. The second term with subscript
``go'' is still calculated within the LO geometric optical WKB approximation
and contains the NTLO-SPA.
Physically, it amounts to displacing the point where specular reflection takes
place from the point of closest approach between the two surfaces. Note that taking
the diffraction contribution (which is already a NTLO term) into account
within the NTLO-SPA would lead to a higher order contribution which can be
neglected here.

In correspondence with the zero-temperature results derived in Ref.~\cite{Henning2019},
the different NTLO contributions obtained from the expansion of (\ref{eq:trMr}) for an
individual Matsubara frequency are given by
\begin{align}
\label{eq:ntlo-d-te}
	[\trace\mathcal{M}(\xi_n)^r]_\mathrm{d}^\mathrm{TE} &= \frac{1}{8}\big[(u^2-4)E_1(u) - (u-1)e^{-u}\big]\\
\label{eq:ntlo-d-tm}
	[\trace\mathcal{M}(\xi_n)^r]_\mathrm{d}^\mathrm{TM} &= -\frac{1}{8}\big[u^2E_1(u) - (u-1)e^{-u}\big]\\
\label{eq:ntlo-go}
	[\trace\mathcal{M}(\xi_n)^r]_\mathrm{go}^p &= -\frac{(r^2-1)e^{-u}}{12r^2}\,, \,\,p=\mathrm{TE,TM}\,.
\end{align}
Here, $E_1$ denotes the exponential integral function~\cite{DLMF} and $u=2Lr\xi_n/c$.
The two polarizations provide identical contributions to the 
 geometrical optics term.

After inserting \eqref{eq:ntlo-d-te}--\eqref{eq:ntlo-go} into the contribution
(\ref{eq:energy_round_trips}) of an individual Matsubara frequency to the
Casimir free energy, we sum over multiple round-trips to find the NTLO
contribution for any non-zero Matsubara frequency
\begin{equation}
\label{sum-n1}
	[\mathcal{F}_n]_\mathrm{\scriptscriptstyle {NTLO}} = 
		 [\mathcal{F}_n]_{\mathrm{d}}^\mathrm{TE}+ [\mathcal{F}_n]_{\mathrm{d}}^\mathrm{TM}
		 	+ [\mathcal{F}_n]_{\mathrm{go}}\,\;\;\;\;(n\neq 0)
\end{equation}
consisting of the contributions from diffraction
\begin{align}
\label{eq:diff_te_correction_sum}
 [\mathcal{F}_n]_\mathrm{d}^\mathrm{TE}  &= -[\mathcal{F}_n]_{\mathrm{d}}^\mathrm{TM}-\frac{\hbar c}{2\lambda_T}
		 	\int_1^\infty\mathrm{d}t \,\frac{\log(1-e^{-4\pi \tau nt})}{t} \\
\label{eq:diff_tm_correction_sum}
[\mathcal{F}_n]_\mathrm{d}^\mathrm{TM}  &=  \frac{\hbar c}{8\lambda_T}\Bigg[(4\pi \tau n)^2\int_1^\infty\mathrm{d}t\frac{e^{-4\pi \tau nt}}{t(1-e^{-4\pi \tau nt})^2} 
- 4\pi \tau n \frac{ e^{- 4\pi \tau n  }}{1- e^{- 4\pi \tau n  }} - \log(1-e^{-4\pi \tau n})\Bigg]
\end{align}
and from geometrical optics
\begin{equation}
\label{eq:go_correction_sum}
 [\mathcal{F}_n]_\mathrm{go}^p = -\frac{\hbar c}{12\lambda_T}
 		\Bigl[\text{Li}_3\left( e^{- 4\pi \tau n  }\right) 
                  +  \log \left(1- e^{- 4\pi \tau n  }\right)\Bigr]\,, \,\,p=\mathrm{TE,TM}\,,
\end{equation}
where $\text{Li}_3$ denotes the trilogarithm~\cite{DLMF}. These results are
valid for arbitrary values of the ratio $\tau= L/\lambda_T$ as long as $R\gg L,
\lambda_T$. In Sec.~\ref{sec:asym_expansion}, when considering the case of
intermediate temperatures $R\gg \lambda_T \gg L$, we will expand
(\ref{eq:diff_te_correction_sum})--(\ref{eq:go_correction_sum}) for $\tau \ll
1$.

\subsection{Zero Matsubara frequency}
\label{sec:Matsubara_n0}

We now turn to the zero-frequency contribution $\mathcal{F}_0$ to the
Casimir free energy and determine the corrections to the PFA result.
At vanishing frequency, the reflection matrix elements of the sphere are
diagonal with respect to polarization \cite{Spreng2018}. For the TM
contribution, the plane-wave approach allows for the derivation of an exact
analytic expression in the more general case of two spheres of arbitrary
radii~\cite{Schoger2020}. The previously known result for the plane-sphere
geometry \cite{Bimonte2012} is recovered as a particular case. The leading
order PFA correction is then found to be proportional to $\log(L/R)$
\begin{equation}\label{eq:asymptotics_TM}
\mathcal{F}^\mathrm{TM}_0 \simeq -\frac{k_\mathrm{B} T}{4}\left[\frac{\zeta(3)}{x} - \frac{1}{6}\log(x) + o(\log(x))\right]\,,
\end{equation}
where we have introduced the dimensionless quantity $x=L/R$ and
$\zeta(3)\approx 1.202$ denotes a particular value of the Riemann zeta
function~\cite{DLMF}.  In the remaining part of this section, we focus on the
asymptotic expansion of the TE contribution to $\mathcal{F}_0$ when the sphere
radius $R$ becomes large compared to the surface-to-surface distance $L$.

The low-frequency limit of the reflection operator at the sphere has been
derived in \cite{Spreng2018}. With Eqs. (8), (A9) and (B6) of
Ref.~\cite{Spreng2018}, the matrix elements for TE polarization read
\begin{equation}\label{eq:TE_reflection_matrix_elements}
\langle \mathbf{k}_{j+1}, -, \mathrm{TE} \vert \mathcal{R}_\mathrm{S} \vert \mathbf{k}_j, +, \mathrm{TE} \rangle = \frac{2\pi R}{k_{j+1}} \sum_{\ell=1}^\infty \frac{\ell}{\ell+1}\frac{y_{j+1,j}^{2\ell}}{(2\ell)!} 
\end{equation}
with
\begin{equation}
y_{j+1,j}=R \sqrt{2(\mathbf{k}_{j+1}\cdot \mathbf{k}_{j} + k_{j+1}k_{j})}\,.
\end{equation}
For large spheres, for which all $y_{j+1,j} \gg 1$, the asymptotics of the reflection
matrix elements \eqref{eq:TE_reflection_matrix_elements} can be obtained by
replacing the sum over $\ell$ by an integral and using Stirling's approximation
for the factorial. The asymptotics of the integral over $\ell$ can then be
found by the leading-order saddle-point approximation with a saddle point at
$\ell_\mathrm{sp} = y_{j+1,j}/2$.  For the asymptotics of the reflection matrix elements
\eqref{eq:TE_reflection_matrix_elements}, we then find
\begin{equation}\label{eq:TE_refl_mat_el_asymptotics}
\langle \mathbf{k}_{j+1}, -, \mathrm{TE} \vert \mathcal{R}_\mathrm{S} \vert \mathbf{k}_j, +, \mathrm{TE} \rangle
= \frac{\pi R}{k_{j+1}}\frac{y_{j+1,j}}{y_{j+1,j} + 2} e^{y_{j+1,j}}
	\left(1+\mathcal{O}\left(\frac{1}{R^2}\right)\right)\,.
\end{equation}
Formally, the zero-frequency limit of \eqref{eq:SA_with_corrections} could be
reproduced by expanding the second factor. However, we need to keep the full
expression to avoid a divergence in the integrals \eqref{eq:LOSP} and
\eqref{eq:ntlo-spa_TE} below. When applying the saddle-point approximation to \eqref{eq:rloops}, the function \eqref{eq:g_definition} thus has
to be replaced by
\begin{equation}
 \label{eq:function_g}
 g(\mathbf{k}_0,\dots,\mathbf{k}_{r-1}) = \prod_{j=0}^{r-1} 
 \frac{e^{-2 k_jL}}{k_j} \frac{y_{j+1,j}}{y_{j+1,j} + 2}
\end{equation}
while in \eqref{eq:f_definition} it is sufficent to set $\xi_n=0$.

We now evaluate the contributions to \eqref{eq:asym_formula} due to
LO-SPA and NTLO-SPA separately. For the LO-SPA of the trace over $r$ round trips, we then find
\begin{equation}\label{eq:LOSP}
[\trace\mathcal{M}(0)^r]^\mathrm{TE}_\mathrm{\scriptscriptstyle {LO-SPA}} \simeq \frac{1}{2r x}\int_0^\infty \mathrm{d}t \left(\frac{t}{t+x}\right)^r e^{-2rt}\,,
\end{equation}
where we have substituted $t=k_\mathrm{sp}L$. Since the sphere radius is
much larger than the distance between plane and sphere, we can 
approximate the integrand in \eqref{eq:LOSP} for $x \ll 1$ and write
\begin{align}
[\trace\mathcal{M}(0)^r]^\mathrm{TE}_\mathrm{\scriptscriptstyle {LO-SPA}} &\simeq \frac{1}{2r x} \int_0^\infty \mathrm{d}t \exp\left[-r(2t + x/t)\right]\nonumber \\
&= \frac{1}{r\sqrt{2 x}} K_1(2r\sqrt{2x})
\label{eq:M_TE_r}
\end{align}
in terms of the modified Bessel function of the second kind $K_1$ \cite{DLMF}.
The terms neglected here contribute to higher order in the asymptotic
expansion.

In view of \eqref{eq:energy_round_trips} we need to evaluate the sum over the number $r$
of round trips of \eqref{eq:M_TE_r} weighted with an additional factor $1/r$. The presence
of the Bessel function leads us to employ a method based on the Mellin transformation \cite{Paris2018}.
The round-trip sum can then be expressed as an integral
\begin{equation}
\sum_{r=1}^\infty\frac{K_1(2r\sqrt{2x})}{r^2} =
\frac{1}{8\pi\text{i}}\int_{c-\text{i}\infty}^{c+\text{i}\infty}\text{d}s
\,\Gamma\left(\frac{s-1}{2}\right)\Gamma\left(\frac{s-3}{2}\right)\zeta(s)(2x)^{-s/2}
\end{equation}
where $\Gamma(z)$ is the Gamma function \cite{DLMF} and the integration contour has to be chosen
such that $c>3$. The integrand contains a single pole at $s=3$, a triple pole at $s=1$ and
double poles at $s=-2n+1$ with $n=1, 2,\ldots$ Keeping only the pole at $s=3$ is equivalent
to PFA and the logarithmic corrections which we are interested in arise from the pole at $s=1$.
Evaluating the corresponding residues and neglecting terms of order one and higher, we
find the asymptotic expansion of the free energy due to the LO-SPA as
\begin{equation}
\left[\mathcal{F}_0^\mathrm{TE}\right]_\mathrm{\scriptscriptstyle {LO-SPA}} \simeq -\frac{k_\mathrm{B} T}{4}\bigg[\frac{\zeta(3)}{x} - \frac{1}{2}\log^2(x) + (1-\log(2))\log(x) + \mathcal{O}(1)\bigg]\,.
\end{equation}

For the NTLO-SPA, we need to evaluate \eqref{eq:F1}. It turns out that we can
partly use the results obtained for finite frequencies in
Ref.~\cite{Henning2019}. Thus, the expressions for the derivatives of the
function $f$ are obtained from (A13) and (A14) of Ref.~\cite{Henning2019} by
taking $\xi\rightarrow 0$. One can show that the contributions to $F_1$ arising
from the derivatives of $f$ cancel out.

In the remaining term in \eqref{eq:F1}, the function $g$ defined in
\eqref{eq:function_g} is differentiated with respect to $v_i$ and $v_{r-i}$.
This term can be decomposed into two contributions,
\begin{equation}
F_1 = \frac{D_{3,1}+D_{3,2}}{2}\,.
\end{equation}
$D_{3,1}$ and $D_{3,2}$ correspond to double derivatives of the two factors
$\exp(-2 k_jL)/k_j$ and $y_{j+1,j}/(y_{j+1,j}+2),$ respectively. The
contribution where a single derivative is taken of each of those factors
vanishes.

The term $D_{3,1}$ can be obtained from (A15) of Ref.~\cite{Henning2019} by taking the zero-frequency limit.
We then find
\begin{equation}
D_{3,1} = - \frac{(r^2-1) L}{3} g\vert_\mathrm{sp}\,.
\end{equation}
In order to determine the term $D_{3,2}$, we follow the procedure described in
Appendix A of Ref.~\cite{Henning2019} and find
\begin{equation}
D_{3,2} =- \frac{(r-1)(3 + (r+1)k_\mathrm{sp} R)}{6 k_\mathrm{sp} (1+k_\mathrm{sp} R)^2}g\vert_\mathrm{sp}\,.
\end{equation}

The NTLO-SPA of the trace over $r$ round trips can then be expressed as
\begin{equation}
\label{eq:ntlo-spa_TE}
[\trace\mathcal{M}(0)^r]^\mathrm{TE}_\mathrm{\scriptscriptstyle {NTLO-SPA}}
= - \frac{r-1}{12 r} \int_0^\infty \mathrm{d}t \left[r+1 + x \frac{3x + (r+1)t}{2t(x+t)^2}\right]\left(\frac{t}{t+x}\right)^r e^{-2 r t}\,,
\end{equation}
where we again used the substitution $t=k_\mathrm{sp}L$. 
For $x\ll 1$, we can write
\begin{equation}
\begin{aligned}
\left[\trace\mathcal{M}(0)^r\right]^\mathrm{TE}_\mathrm{\scriptscriptstyle {NTLO-SPA}} &\simeq - \frac{r^2-1}{12 r} \int_0^\infty \mathrm{d}t \left[1 +  \frac{x}{2 t^2}\right] e^{-r (2t+x/t)}\\
&= -\frac{r^2-1}{6r} \sqrt{2x}\, K_1(2r\sqrt{2x}) \,.
\end{aligned}
\end{equation}
Note that the second term in the square bracket above needs to be kept, as it is asymptotically of the same order as the first one.
Performing the sum over round trips using the method of Ref.~\cite{Paris2018}, we find the NTLO saddle-point contribution to the Casimir free energy for TE polarization as 
\begin{equation}\label{eq:asymptotics_TE}
\left[\mathcal{F}_0^\mathrm{TE}\right]_\mathrm{\scriptscriptstyle {NTLO-SPA}} \simeq - \frac{k_\mathrm{B} T}{24} \log(x)\,.
\end{equation}
The total TE contribution to the free energy then becomes
\begin{equation}\label{eq:F_0_TE_asymptotics}
\begin{aligned}
\mathcal{F}_0^\mathrm{TE} & \simeq \left[\mathcal{F}_0^\mathrm{TE}\right]_\mathrm{\scriptscriptstyle {LO-SPA}} + \left[\mathcal{F}_0^\mathrm{TE}\right]_\mathrm{\scriptscriptstyle {NTLO-SPA}} \\
 &= -\frac{k_\mathrm{B} T}{4}\bigg[\frac{\zeta(3)}{x} - \frac{1}{2}\log^2(x)
 + \left(\frac{7}{6}-\log(2)\right)\log(x) + o(\log(x))\bigg]\,.
\end{aligned}
\end{equation}

\begin{figure}
\centering
\includegraphics{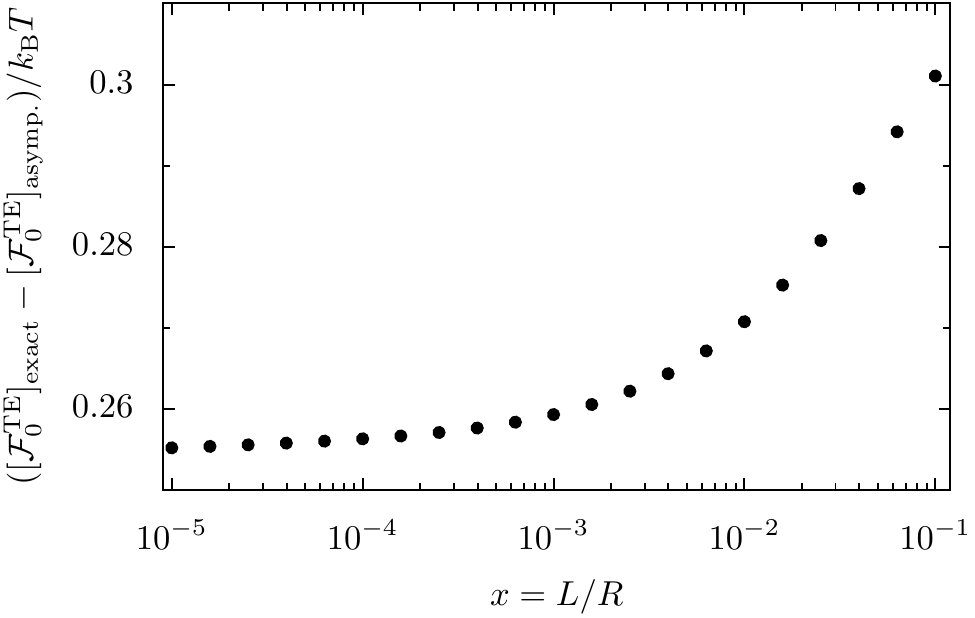}
\caption{Numerical analysis of the correction to formula \eqref{eq:F_0_TE_asymptotics}.
The dots show the difference between the numerically exact values of the
zero-frequency contribution to the Casimir free energy for TE polarization
and the corresponding values according to formula \eqref{eq:F_0_TE_asymptotics}.
This difference is shown as a function of $x=L/R.$}
\label{fig:nntlo_correction}
\end{figure}

To verify that the asymptotic expression for the zero-frequency contribution to
the Casimir free energy due to TE polarization given by
\eqref{eq:F_0_TE_asymptotics} is correct, we compare with the corresponding
numerically exact result. In Fig.~\ref{fig:nntlo_correction}, the difference
between the asymptotics and the numerical exact result is shown as a function
of $x=L/R$.  As this difference is decreasing with decreasing values of $x$ and
thus subleading compared to $\log(x)$, our numerical comparison shows that the
asymptotic expansion \eqref{eq:F_0_TE_asymptotics} is indeed correct.

The complete asymptotic expansion for the zero-frequency contribution is obtained
by adding the contributions of the TM polarization \eqref{eq:asymptotics_TM} and
the TE polarization \eqref{eq:F_0_TE_asymptotics} and will explicitly be given
and used in the next section in \eqref{0-matsubara}.

\section{Leading-order correction to PFA at intermediate temperatures }
\label{sec:asym_expansion}

The contribution of thermal fluctuations to the leading-order correction to the
PFA result for the Casimir free energy is derived as the difference between the
Matsubara sum of $[\mathcal{F}_n]_\mathrm{\scriptscriptstyle {NTLO}}$ and the
corresponding integral representing the zero-temperature limit. Such a
difference is usually evaluated with the help of the
Abel-Plana~\cite{Bordag2010} or the Poisson summation~\cite{Ingold2009}
formula. Since the zero-frequency contribution was treated separately in the
previous section, we found it more convenient to make use of the
Euler-Maclaurin formula in the form
\begin{equation}
\label{euler-maclaurin}
\sum_{n=1}^\infty \mathcal{F}_n = \int_1^\infty\mathrm{d} n\, \mathcal{F}_n + \frac{\mathcal{F}_1 + \mathcal{F}_\infty}{2} +
\sum_{m=1}^{\infty}\frac{B_{2m} \left[\mathcal{F}^{(2m-1)}_\infty - \mathcal{F}^{(2m-1)}_1\right]}{(2m)!}\,,
\end{equation}
where $\mathcal{F}^{(2m-1)}_n$ denotes the $(2m-1)$-th derivative of
$\mathcal{F}_n$ with respect to $n$ taken as a continuous variable and $B_{2m}$
are the Bernoulli numbers \cite{DLMF}.

We apply the Euler-Maclaurin formula (\ref{euler-maclaurin}) to the sum of the
NTLO Matsubara contributions $[\mathcal{F}_n]_\mathrm{\scriptscriptstyle
{NTLO}}$ given by Eqs.~(\ref{sum-n1})--(\ref{eq:go_correction_sum}). These
terms decay for large frequencies so that
$[\mathcal{F}_\infty]_\mathrm{\scriptscriptstyle {NTLO}}$ and all its
derivatives vanish. Only the first Matsubara frequency contributes to the
second and third terms in (\ref{euler-maclaurin}). For $\tau = L/\lambda_T \ll1$, the leading
contribution to $[\mathcal{F}_1]_\mathrm{\scriptscriptstyle {NTLO}}$ arises
from the TE diffraction term (\ref{eq:diff_te_correction_sum})
\begin{equation}
\label{eq: diff_te_corr_int2}
 [\mathcal{F}_1]_{\mathrm{d}}^\mathrm{TE} \simeq  \frac{\hbar c\tau}{4L}\log^2(\tau)\,.
\end{equation}
Relative to the zero-temperature NTLO result, the TE diffraction contribution
is $\mathcal{O}(\tau\log^2(\tau)),$ whereas both the TM diffractive
contribution (\ref{eq:diff_tm_correction_sum}) and the geometric optical
contributions (\ref{eq:go_correction_sum}) are $\mathcal{O}(\tau\log \tau)$.
In addition, the derivatives of $[\mathcal{F}_1]_\mathrm{\scriptscriptstyle
{NTLO}}$ appearing in the Euler-Maclaurin formula (\ref{euler-maclaurin}) are
also $\mathcal{O}(\tau\log \tau)$ and hence can be neglected. 

In order to connect the integral appearing on the right-hand side of
(\ref{euler-maclaurin}) with the zero-temperature result, we need to
account for the difference in the lower bound. Thus, we obtain the NTLO
terms from (\ref{euler-maclaurin}) as 
\begin{equation}
\label{euler-maclaurin2}
\sum_{n=1}^{\infty} [\mathcal{F}_n]_\mathrm{\scriptscriptstyle {NTLO}} \simeq
 [\mathcal{F}(T=0)]_\mathrm{\scriptscriptstyle {NTLO}}-
 \int_0^1\mathrm{d} n\,  [\mathcal{F}_n]_\mathrm{\scriptscriptstyle {NTLO}} + \frac{ [\mathcal{F}_1]_{\mathrm{d}}^\mathrm{TE}}{2}
\end{equation}
where $[\mathcal{F}(T=0)]_\mathrm{\scriptscriptstyle {NTLO}}$ denotes the NTLO
contribution to the free energy in the zero-temperature limit.  The integral
subtracted on the right-hand side of (\ref{euler-maclaurin2}) is of the same
order as $[\mathcal{F}_1]_{\mathrm{d}}^\mathrm{TE}$ and its leading-order
contribution also arises from the TE diffraction term
(\ref{eq:diff_te_correction_sum}). We find
\begin{equation}
\label{eq:n_pos_cont}
	\sum_{n=1}^{\infty}[\mathcal{F}_n]_\mathrm{\scriptscriptstyle {NTLO}} 
		 \simeq  [\mathcal{F}(T=0)]_\mathrm{\scriptscriptstyle {NTLO}} - \frac{\hbar c \tau}{8 L}\log^2(\tau)\,.
\end{equation}

Finally, it is still necessary to add the Matsubara zero-frequency contribution
$[\mathcal{F}_0]_\mathrm{\scriptscriptstyle {NTLO}} $ to (\ref{eq:n_pos_cont})
in order to obtain the full NTLO Casimir free energy from
(\ref{eq:Matsubara_sum}).  Naively, one could expect that
$[\mathcal{F}_0]_\mathrm{\scriptscriptstyle {NTLO}}$ would not contribute in
the limit $\tau\ll 1$. However, this term is relevant for intermediate
temperatures $L/R \ll\tau\ll 1$ as far as the correction to PFA is
concerned. In Sec.~\ref{sec:Matsubara_n0}, we have found that the zero-frequency
contribution to the Casimir free energy with \eqref{eq:asymptotics_TM} and
\eqref{eq:F_0_TE_asymptotics} reads, up to NNTLO, 
\begin{equation}\label{0-matsubara}
\mathcal{F}_0 \simeq -\frac{\hbar c\tau}{4L}\bigg[\frac{2\zeta(3)}{x} - \frac{1}{2}\log^2(x) + \left(1-\log(2)\right)\log(x) + O(1)\bigg]\,,
\end{equation}
where $x=L/R$ was introduced at the beginning of Sec.~\ref{sec:Matsubara_n0}. The NTLO and
NNTLO zero-frequency contributions correspond to the second and third terms on
the right-hand side of \eqref{0-matsubara}.  They are both asymptotically
larger than the thermal correction arising from nonzero frequencies given by
(\ref{eq:n_pos_cont}) when $x\ll \tau\ll 1$. In practice, however, all
those contributions are comparable in the case of experimentally relevant
values of $\tau$ and $x,$ as illustrated by the numerical example discussed
below. 

As we want to focus on the interplay between geometrical and thermal effects,
we first define the total thermal correction to the Casimir free energy
\begin{equation}
\label{eq:delta_F}
\delta \mathcal{F}(T) =  \mathcal{F}(T) -  \mathcal{F}(T=0).
\end{equation}
and introduce the deviation of the thermal correction from the PFA result
relative to the zero-temperature PFA free energy   
\begin{equation}
 \label{eq:Delta}
 \Delta = \frac{\delta\mathcal{F}(T)-\delta\mathcal{F}_\mathrm{PFA}(T)}
	       {\mathcal{F}_\mathrm{PFA}(T=0)}\,.
\end{equation}
After taking \eqref{eq:n_pos_cont} and \eqref{0-matsubara} into account, we find
for intermediate temperatures $x\ll\tau\ll 1$
\begin{equation}
\label{eq:full-expression}
 \Delta \simeq \frac{45}{\pi^3}x\tau\left[-\log^2(x)+2[1-\log(2)]\log(x)
	 +2\log^2(\tau)+\mathcal{O}\big(\log(\tau)\big)\right]\,,
\end{equation}
where the leading neglected terms arise from non-zero Matsubara frequencies.

\begin{figure}
\centering
\includegraphics[width=0.8\columnwidth]{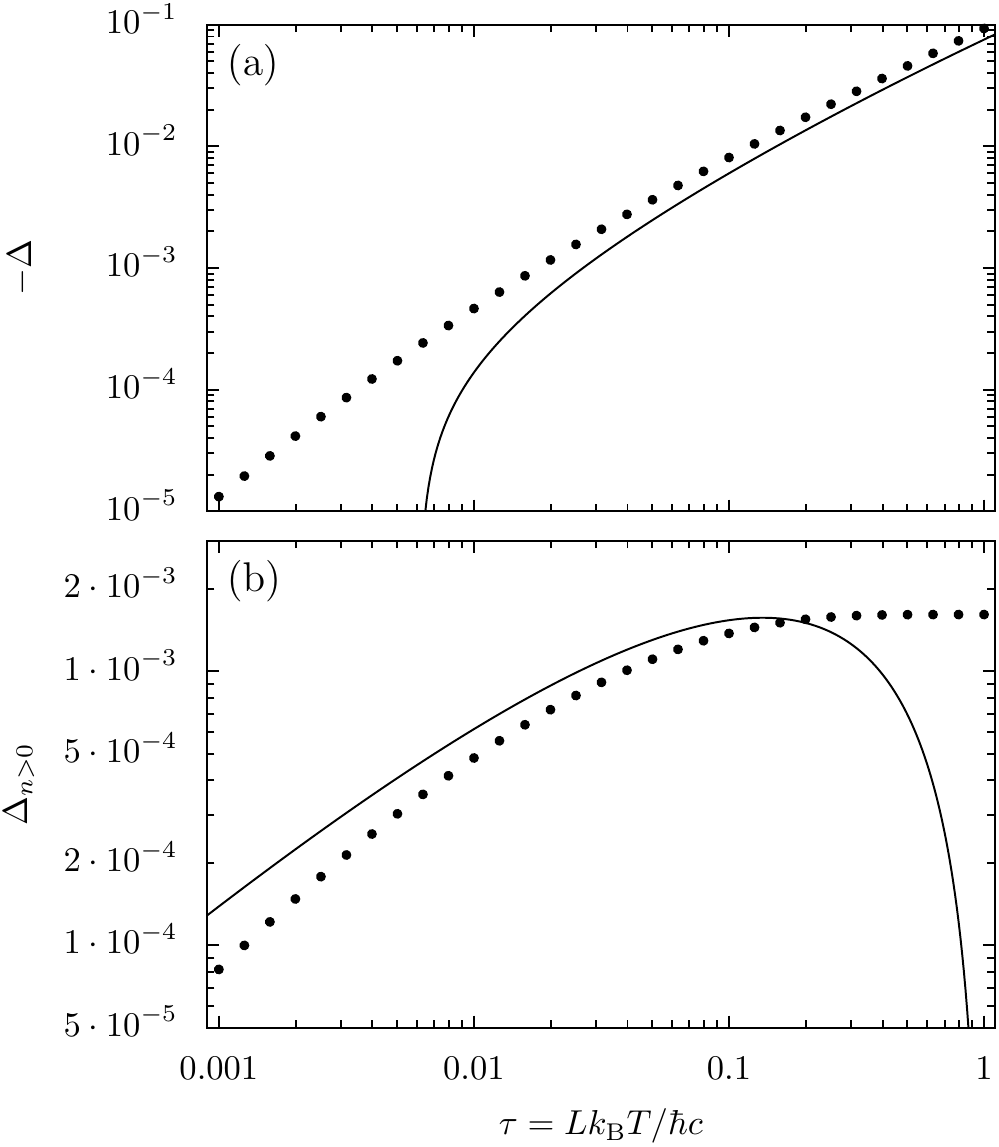}
\caption{Relative thermal correction \eqref{eq:Delta} of the Casimir free energy
as a function of temperature for a geometrical aspect ratio of $x=L/R=10^{-3}$.
(a) The dots represent the numerically exact full Matsubara sum and the line
corresponds to the analytical asymptotic expansion \eqref{eq:full-expression}.
Note that the correction $\Delta$ is negative. (b) Contribution $\Delta_{n>0}$
of the positive Matsubara frequencies $n>0$. The dots again represent the numerically
exact result while the line corresponds to the analytical asymptotic expansion derived
from \eqref{eq:n_pos_cont}. For $n>0$, the correction to the PFA result is positive.}
\label{lowT_R1000}
\end{figure}

In Fig.~\ref{lowT_R1000}, we show the correction $\Delta$ as a function of
temperature. The geometrical aspect ratio is chosen as $x=L/R=10^{-3}$, a
typical order-of-magnitude in most Casimir experiments~\cite{Hartmann2018}.  In
the upper panel (Fig.~\ref{lowT_R1000}a), the full Matsubara sum is considered.
The dots represent the exact correction as calculated by the numerical method
presented in Ref.~\cite{Spreng2020}, whereas the line corresponds to the
analytical approximation (\ref{eq:full-expression}). In contrast, in the lower
panel (Fig.~\ref{lowT_R1000}b), the contribution of the zero Matsubara
frequency has been disregarded. Note that the sign of $\Delta$ in the two
panels differs.

According to the results displayed in Fig.~\ref{lowT_R1000}b, the sum over
nonzero frequencies is well described by the analytical formula derived from
(\ref{eq:n_pos_cont}) in the range of intermediate temperatures $x\ll\tau\ll
1.$ The resulting correction is positive, thus reducing the total correction to
PFA. In contrast, the total thermal contribution to the PFA correction, with the
zero-frequency contribution included, is negative in the entire range shown in
the figure. Thus, the strength of the interaction is further reduced with
respect to the PFA prediction due to thermal effects. The zero-temperature
result underestimates the total correction to PFA by a factor of about two at
$\tau\approx 3\times 10^{-2},$ indicating the strong interplay between thermal
and geometrical effects~\cite{Weber2010,Weber2010B}. 

The zero-frequency contribution plays a significant role in such interplay, as
the total PFA correction and the thermal contribution from nonzero frequencies
have opposite signs in the entire range shown in Fig.~\ref{lowT_R1000}a.  We
find good agreement between the data and our analytical formula
\eqref{eq:full-expression} for $x\ll \tau.$ Since the zero-frequency
contribution becomes increasingly dominant as the temperature rises above $\tau
\sim  0.1,$ formula \eqref{eq:full-expression} also provides a good description
even beyond the range of intermediate temperatures.

The results obtained in this section allow for the derivation of the 
NTLO Casimir entropy for intermediate temperatures. 
By adding \eqref{0-matsubara} to \eqref{eq:n_pos_cont}
and neglecting sub-leading contributions when
taking the derivative with respect to temperature, we find
\begin{equation}
\label{entropy-ntlo}
	S_\mathrm{\scriptscriptstyle {NTLO}} \simeq \frac{k_\mathrm{B}}{16}\left[
		2\log^2\left(\tau\right) 
		-\log^2\left(\frac{L}{R}\right) +2(1-\log(2))\log\left(\frac{L}{R}\right)\right]\,.
\end{equation}
The first term on the right-hand side results from the contribution of nonzero
frequencies.  The zero-frequency contribution,  represented by the second and
third terms, corresponds to a temperature-independent, negative contribution
reminiscent of the negative Casimir entropies found for aspect ratios $L/R\sim
1$ or larger \cite{Canaguier-Durand2010B,Umrath2016}.

\section{Conclusion}
\label{sec:conclusions}

We have analyzed the leading-order correction to PFA in the plane-sphere
geometry for intermediate temperatures satisfying the condition $x=L/R\ll \tau
= L k_\mathrm{B}T/\hbar c \ll 1$, which holds in most Casimir force experiments.
Whereas the Matsubara zero frequency is unimportant for extremely low
temperatures satisfying $\tau\ll x\ll 1$, it provides a sizeable contribution
to the correction in the case of intermediate temperatures. When considering
its asymptotic limit for $R\gg L$, we should keep not only Bimonte's NTLO
term~\cite{Bimonte2017}, proportional to $x\tau\log^2(x)$, but also the NNTLO
term proportional to $x\tau\log(x)$ in order to have an accurate formula for
experimentally-relevant aspect ratios.  We have also derived an additional
logarithmic term of the form $x\tau\log^2(\tau)$ by considering the
contribution of nonzero frequencies.  As an effect of the logarithmic terms,
the zero-temperature result grossly underestimates the correction to PFA even
at the rather low temperatures $\tau\sim 10^{-2}$ corresponding to typical
experiments.  Altogether our findings demonstrate the strong interplay between
thermal and beyond-PFA geometrical corrections.

\vspace{6pt}

\supplementary{The data represented in Figures 2 and 3 are freely available from Zenodo at
\url{https://doi.org/10.5281/zenodo.4631940}. \cite{Zenodo2021}}

\authorcontributions{V. H. and B. S. developed the theoretical formalism and B. S. carried
out the numerical calculations. P. A. M. N. and G.-L. I. coordinated the work. All authors
discussed the results and co-wrote the manuscript.}

\funding{Deutscher Akademischer Austauschdienst (DAAD); Conselho Nacional de
Desenvolvimento Cient\'{\i}fico e Tecnol\'ogico (CNPq), Coordena\c c\~ao de
Aperfei\c coamento de Pessoal de N\'{\i}vel Superior (CAPES); Instituto Nacional
de Ci\^encia e Tecnologia Fluidos Complexos (INCT-FCx); Funda\c c\~ao Carlos
Chagas Filho de Amparo \`a Pesquisa do Estado do Rio de Janeiro (FAPERJ);
Funda\c c\~ao de Amparo \`a Pesquisa do Estado de S\~ao Paulo (FAPESP).}

\acknowledgments{The authors would like to thank Michael Hartmann and Tanja Schoger
for stimulating discussions. This work has been supported by CAPES and DAAD
through the PROBRAL collaboration program.}

\conflictsofinterest{The authors declare no conflict of interest.}

\abbreviations{The following abbreviations are used in this manuscript:\\

\noindent
\begin{tabular}{@{}ll}
PFA & proximity-force approximation\\
LO & leading order\\
LO-SPA & leading order saddle-point approximation\\
NTLO & next-to-leading order\\
NTLO-SPA & next-to-leading order saddle-point approximation\\
NNTLO  & next-to-next-to-leading order\\
TE & transverse electric\\
TM & transverse magnetic\\
WKB & Wentzel-Kramers-Brillouin\\
\end{tabular}}

\appendixtitles{yes}
\appendixstart
\appendix
\section{Next-to-leading-order correction in the saddle-point approximation}
\label{app:SP_formula}

In the main part of this paper, we need to asymptotically evaluate an integral of the form
\begin{equation}
 \label{eq:integral1d}
 I = \int\text{d}^d\mathbf{x}\,g(\mathbf{x})\exp\big(-Rf(\mathbf{x})\big)
\end{equation}
for large values $R$ where $\mathbf{x}=(x_1,\ldots,x_d)$ is a $d$-dimensional vector.
To keep the discussion simple, we start with the one-dimensional case and
merely state the result for the multi-dimensional case at the end. Furthermore,
we will assume the existence of only a single saddle point (sp),
\textit{i.e.} a point where the first derivative $f'(x)$ vanishes, and this point should
lie well inside the range of integration. This will be the case in our application.

Using Laplace's method, one obtains the well-known leading order of the
saddle-point approximation (LO-SPA) of the integral \eqref{eq:integral1d} as
\begin{equation}
 \label{eq:lo_spa_1d}
 I_\text{LO-SPA} = \left(\frac{2\pi}{Rf_\text{sp}''}\right)^{1/2}g_\text{sp}\exp(-Rf_\text{sp})\,.
\end{equation}
We assume here that the second derivative $f_\text{sp}''$ at the saddle point
is positive. $f_\text{sp}$ and $g_\text{sp}$ denote the value of the functions
$f(x)$ and $g(x)$, respectively, at the saddle point.

For our purposes, we also need the next-to-leading-order term of the saddle-point
approximation (NTLO-SPA) which relative to the LO-SPA carries an additional
factor $1/R$ and which we will derive now. For a nonvanishing second derivative
$f_\text{sp}''$ only a region of width $R^{-1/2}$ around the saddle point contributes
to the integral \eqref{eq:integral1d}. We therefore extend the Taylor expansion in
the exponent up to fourth order and expand the exponential containing the third
and fourth order terms into a Taylor series. Keeping only terms contributing to the
LO-SPA and the NTLO-SPA after integration, we can approximate the exponential by
\begin{equation}
 \label{eq:f_expansion}
 \begin{aligned}
  \exp\big(-Rf(x)\big) &\approx \exp(-Rf_\text{sp})\exp\left[-\frac{R}{2}f_\text{sp}''x^2\right]\\
     &\quad\times\left(1-\frac{R}{6}f_\text{sp}'''x^3-\frac{R}{24}f_\text{sp}''''x^4
 	+\frac{R^2}{72}f_\text{sp}'''{}^2x^6\right)\,.
 \end{aligned}
\end{equation}
Here, we have assumed for simplicity that the saddle point is located at $x=0$.
In addition, we need to expand the function $g(x)$ up to second order
\begin{equation}
 \label{eq:g_expansion}
 g(x) \approx g_\text{sp}+g_\text{sp}'x+\frac{1}{2}g_\text{sp}''x^2\,.
\end{equation}
Inserting \eqref{eq:f_expansion} and \eqref{eq:g_expansion} into \eqref{eq:integral1d}
for $d=1$, the integration can be carried out and we obtain
\begin{equation}\label{eq:SP_formula}
 I = I_\text{LO-SPA}+\frac{1}{R}I_\text{NTLO-SPA}+\mathcal{O}(R^{-2})
\end{equation}
with
\begin{equation}
 \label{eq:NTLO-SPA_1d}
 I_\text{NTLO-SPA} = I_\text{LO-SPA}\left(\frac{1}{2}\frac{g_\text{sp}''}{g_\text{sp}f_\text{sp}''}
   -\frac{1}{2}\frac{g_\text{sp}'f_\text{sp}'''}{g_\text{sp}f_\text{sp}''{}^2}
   -\frac{1}{8}\frac{f_\text{sp}''''}{f_\text{sp}''{}^2}
   +\frac{5}{24}\frac{f_\text{sp}'''{}^2}{f_\text{sp}''{}^3}\right)\,.
\end{equation}

In the multi-dimensional case, the generalization of the result \eqref{eq:lo_spa_1d}
for the leading order is well-known to read
\begin{equation}\label{eq:LO-SPA_multi}
I_\text{LO-SPA} = \left(\frac{2\pi}{R}\right)^{d/2}  \frac{e^{-R f_\mathrm{sp}}}{\sqrt{\det \mathsf{H}}} g_\mathrm{sp}
\end{equation}
with the Hessian matrix
\begin{equation}
\mathsf{H} \equiv \left(\left.\frac{\partial^2 f}{\partial x_i \partial x_j}\right\vert_\mathrm{sp}\right)_{i,j=1,\ldots,d}
\end{equation}
which is assumed to be non-singular.
Proceeding along the lines explained for the one-dimensional case,
the next-to-leading term in the saddle-point approximation becomes
\begin{multline}
 \label{eq:NTLO-SPA_multi}
 I_\text{NTLO-SPA} = I_\text{LO-SPA}\bigg[\frac{1}{2}\frac{g_{ij}\mathsf{H}^{ij}}{g_\text{sp}}
     - \frac{1}{2}\frac{f_{ijk}g_l\mathsf{H}^{ij}\mathsf{H}^{kl}}{g_\text{sp}}
     - \frac{1}{8}f_{ijkl}\mathsf{H}^{ij}\mathsf{H}^{kl}\\
 +\frac{1}{24} f_{ijk} f_{lmn}\Big(3\mathsf{H}^{ij}\mathsf{H}^{kl}\mathsf{H}^{mn} 
     + 2 \mathsf{H}^{il}\mathsf{H}^{jm}\mathsf{H}^{kn}\Big) \bigg]
\end{multline}
where the subscript ``sp'' denotes the evaluation of the function at
$\mathbf{x}=\mathbf{x}_\mathrm{sp}$. A derivative with respect to the $i$-th
component of $\mathbf{x}$ with subsequent evaluation at the saddle-point is
represented by a lower index $i$: $f_i \equiv \partial f/\partial x_i
\vert_{\mathbf{x}=\mathbf{x}_\mathrm{sp}}$ and equivalently for $g$. Likewise,
higher-order derivatives are denoted by multiple lower indices. Two upper
indices denote the matrix elements of the inverse matrix, $\mathsf{H}^{ij}
\equiv (\mathsf{H}^{-1})_{ij}$, and the Einstein summation convention is 
implied, \textit{i.e.} indices occuring both as sub- and superscript within a term are
summed over with values running from $1$ to $d$. The relation between the result
\eqref{eq:NTLO-SPA_multi} and the one-dimensional result \eqref{eq:NTLO-SPA_1d}
is rather straightforward except for the last two terms in \eqref{eq:NTLO-SPA_multi}.
They account for different index pairings and collapse into a single term in
the one-dimensional case, \textit{i.e.} the last term in \eqref{eq:NTLO-SPA_1d}.

\end{paracol}
\reftitle{References}

\externalbibliography{yes}
\bibliography{casimir}

\end{document}